TorchDIVA: An Extensible Computational Model of Speech Production built on an Open-Source Machine Learning Library

*Sean Kinahan, Julie Liss, Visar Berisha*
Arizona State University
{skinahan, jmliss, visar}@asu.edu

# Abstract

The DIVA model is a computational model of speech motor control that combines a simulation of the brain regions responsible for speech production with a model of the human vocal tract. The model is currently implemented in Matlab Simulink; however, this is less than ideal as most of the development in speech technology research is done in Python. This means there is a wealth of machine learning tools which are freely available in the Python ecosystem that cannot be easily integrated with DIVA. We present TorchDIVA, a full rebuild of DIVA in Python using PyTorch tensors. DIVA source code was directly translated from Matlab to Python, and built-in Simulink signal blocks were implemented from scratch. After implementation, the accuracy of each module was evaluated via systematic block-by-block validation. The TorchDIVA model is shown to produce outputs that closely match those of the original DIVA model, with a negligible difference between the two.

We additionally present an example of the extensibility of TorchDIVA as a research platform. Speech quality enhancement in TorchDIVA is achieved through an integration with an existing PyTorch generative vocoder called DiffWave. A modified DiffWave mel-spectrum upsampler was trained on human speech waveforms and conditioned on the TorchDIVA speech production. The results indicate improved speech quality metrics in the DiffWave-enhanced output as compared to the baseline. This enhancement would have been difficult or impossible to accomplish in the original Matlab implementation. This proof-of-concept demonstrates the value TorchDIVA will bring to the research community. Researchers can download the new implementation at:

https://github.com/skinahan/DIVA_PyTorch

# Introduction

The Directions into Velocities of Articulators (DIVA) model is among the most widely used computational models of the mechanisms responsible for speech motor control [1]. DIVA combines a simulation of the brain regions responsible for speech production with a model of the human vocal tract. Its impact in the speech science community has been considerable [2]; however, we posit that it's limited adoption in the speech technology community is because of its current implementation in Matlab Simulink. While progress in speech technology has been rapid owing to the progress made in large-scale deep learning models [2], one of the rate-limiting factors with applying these tools to emerging applications is a lack of data. For example, while speech recognition datasets are on the order of hundreds of thousands of hours of data [2], clinical speech data sets are many orders of magnitude smaller [2]. One way to overcome the lack of data is to reduce the solution space of these large models by applying constraints from domain knowledge about the data generating process. Theoretically, DIVA can be used for this purpose, however its current implementation makes this a challenge, as most of the existing speech technology tools are developed in Python ML frameworks, while DIVA is built on Matlab Simulink. The current model implementation also makes it difficult to extend DIVA by making use of new speech tools developed by the speech technology community. In this paper, we present a new implementation of the DIVA model in PyTorch, an open-source machine learning framework, validate it relative to the original implementation, and provide a proof-of-concept example of how the DIVA model can be integrated with other tools available in PyTorch to improve the quality of speech it produces.

      DIVA provides a computational model of speech production. Within DIVA, feedforward and feedback control loops are combined with neural networks. These neural networks correspond with neuronal populations in the cerebral cortex, allowing for direct activity comparison against neuroimaging data [1]. The neural network components allow the model to adaptively synthesize speech and compensate for auditory or somatosensory perturbations in real-time. The design of these neural networks is based upon functional magnetic resonance imaging (fMRI) studies on speech acquisition and production [3]. Production of speech in the DIVA model is done via the Maeda method, a time-domain simulation of the vocal tract controlled by articulatory parameters [4]. This combination of

a control model and a mechanical vocal tract model results in a flexible and expressive articulatory speech synthesizer [5]. A simplified view of the overall DIVA model architecture is available in (Fig 1).

**Fig 1. DIVA Model Architecture.** Simplified schematic view of the DIVA model, showing the combination of feedforward and feedback control loops.

Models like DIVA serve to condense and validate the broader speech community's understanding of speech motor control processes. Functional brain imaging studies in the early 2000's enabled speech researchers to identify the brain areas recruited during speech motor control [6]. The brain areas of interest in speech production include those regions commonly associated with planning and execution of movement, as well as centers of acoustic and phonological processing of speech sounds [6]. DIVA allows researchers to investigate the connection between these neurological bases of motor control and their associated model mechanisms [5]. These studies have yielded data of significant clinical value. For example, DIVA has been applied to evaluate the relationship between neurological deficits in auditory and motor processes which underlie pediatric motor speech disorders [7] and to investigate the role of auditory feedback bias in stuttering [8].

Speech targets in the DIVA model are represented by a combination of acoustic and somatosensory target regions. These regions define what the expected acoustic and somatosensory state of the model should be across the time-course of the speech production simulation. The neural networks within the DIVA model are the critical components responsible for transforming the targets defined in acoustic and somatosensory space into motor movements in articulatory space.

The neural network components of DIVA consist of synaptic weight elements used to transform the signal from one neural representation to another. Standard weight elements are used in the speech sound map and error map modules. Standard weights are static and dependent upon the speech target definition. Inverse-map elements take auditory or somatosensory feedback as input and compute an error signal via a Jacobian pseudoinverse mapping of the vocal tract model. Auditory and somatosensory error signals are combined to generate the feedback motor command. The inverse-map elements are static during normal DIVA operation, however they are trained during a babbling phase of random speech articulator movements [6]. The adaptive weight element is a critical submodule of the motor cortex module. This adaptive weight block is responsible for generating the feedforward motor command using synaptic weights learned over multiple iterations of the model.

The long-term goal of the DIVA model is to construct "a comprehensive, unified account of the neural mechanisms underlying speech motor control" [6, p. 13]. We posit that advancing the current incarnation of the model can help spur progress towards this long-term goal. The publicly available implementation of DIVA is built in Matlab Simulink. Simulink is a model-based design and simulation tool, primarily used in signal processing and hardware simulation workflows. It works well for the design and modeling of modular signal-processing based tools; however, DIVA is largely based on neural networks. While training neural networks in Matlab is supported, most speech technology tools are developed in Python ML frameworks, in particular PyTorch [9-13]. As Matlab does not support direct import of PyTorch models, this limits the extensibility of the current DIVA model.

The last several years have seen an exponential increase in the popularity of neural networks (e.g. deep learning) in academic fields. This has led to an explosion in the tools available for training networks and performing inference on them. Significant development in this area has been done in Python open-source machine learning frameworks (e.g. PyTorch, TensorFlow) [14]. These trends have resulted in many sophisticated open-source tools for processing speech and speech audio. Some examples of these tools are pyAudioAnalysis [15], PyTorch-Kaldi [10], SpeechBrain [11], ASVtorch [9], WaveNet [16], and DiffWave [12]. The current DIVA implementation in Simulink does not integrate directly with these tools and deep learning frameworks. In this paper we present TorchDIVA, an implementation of the DIVA model in its entirety using the PyTorch framework.

PyTorch is an open-source machine learning framework with built-in support for critical machine learning features such as backpropagation and scalable, distributed training [17]. Additionally, PyTorch readily integrates with a host of scientific computing, visualization and analysis packages commonly used by deep learning researchers such as NumPy, Seaborn, and SciPy. These features, along with the familiar syntax (to those that know Python) of the framework, will grant greater ease-of-use and flexibility to speech researchers. By boosting the utility and accessibility of DIVA, we hope to enable more researchers, including those with machine learning backgrounds, to contribute to the continued development of the DIVA model.

TorchDIVA is useful from a speech science research perspective because it simplifies the model refinement process. This new implementation enables neural network models of DIVA's brain regions to be easily updated and replaced to incorporate the latest neuroimaging findings. DIVA has repeatedly proven to be a valuable representation of the neural mechanisms of speech production and acquisition [1]. TorchDIVA directly benefits speech technology researchers by enabling the combination of the DIVA speech production model with existing machine learning pipelines. Constraining machine learning models with the domain expertise built into TorchDIVA has the potential to reduce the sample size requirements during training. A reduction in sample size is especially useful for clinical applications, where data is typically scarce.

In this paper, we validate the TorchDIVA implementation module-by-module relative to the original DIVA implementation in Matlab Simulink. In addition, to provide a proof-of-concept integration between an existing model built in PyTorch (DiffWave [12]) and TorchDIVA to produce higher quality speech output. Despite the sophistication of the DIVA architecture, it lacks direct integration with human speech experiments. New speech targets in the Matlab Simulink implementation of DIVA must be defined by hand in a specific input format. Speech generated by DIVA and TorchDIVA is also recognizably artificial to a human listener. The quality of the speech audio is generally low, and the output audio lacks the acoustic characteristics of real human speech. We extended TorchDIVA to enable the automated generation of speech based on samples from a human speaker as target.

## Method

The DIVA model architecture can be divided into several primary modules and sub-modules. These modules have been individually rebuilt in Python, using PyTorch tensors to carry signals and perform operations. Modules consisting of user-defined code (e.g. S2 Matlab functions) were directly translated from Matlab code to Python. Built-in Simulink signal blocks with no directly available source code (e.g. discrete delay, FIR filter) were implemented from scratch in Python. After implementation, the accuracy of each module was evaluated via systematic block-by-block validation. We use several predefined speech targets as test cases. These predefined targets were the built-in cases provided with the original implementation of DIVA. The targets include the phonemes 'i', 'u', 'e', and 'ae', and the word 'happy'. Then we gather recordings from the original DIVA model representing signal input-output pairs of a module during the model activation timeseries for each target. These signal pairs were exported from the Matlab environment and used to craft basic automated tests for each module of TorchDIVA. The tests ensure that each module meets its design and behaves as intended. Once these modules were validated, compound behaviors between modules were similarly evaluated to create comprehensive tests of the model.

## Module Validation

Each TorchDIVA module has been validated by testing against the corresponding DIVA module as a baseline. Nearly all modules of the DIVA model are deterministic, meaning that they repeatably yield the same output when presented with the same input. For the motor cortex module specifically, we evaluate the normalized root mean squared error (RMSE) between the motor command signal output and the equivalent signal in TorchDIVA. The normalized RMSE is calculated by representing the absolute signal difference as a percentage of the motor command's maximum possible amplitude. This method was applied to all modules implemented during this phase.

## Motor Cortex

The motor cortex submodule contains the neural network components responsible for acquiring the motor programs for speech sounds. In the trained model, these motor programs are used to generate sequences of feedforward motor commands which correspond to the learned speech sounds [1]. The learning process for a speech target can be simulated in the DIVA model by resetting the feedforward synaptic weights. This forces the DIVA model to reacquire a motor program for the current speech target using the auditory and somatosensory feedback provided by the feedback control loop. Acquisition of the feedforward motor program was validated by resetting the feedforward synaptic weights in both models, and then re-learning these over the course of 20 iterations. This process was performed for each of the predefined speech targets defined in the original DIVA model. These tests were used to verify the correct functioning of the adaptive feedback control system in the motor cortex module of TorchDIVA. As with previous validation tests, this procedure was repeated for each of the five built-in speech targets included with the Matlab DIVA implementation. This iterative process was also repeated for the well-learned speech targets, without resetting the forward mapping. In each case, the motor commands emitted by the motor cortex modules were recorded, and a normalized RMSE was calculated for each sequence pair.

## Vocal Tract

The DIVA model is largely deterministic and computationally explicit [6]. Therefore, the testing methods applied assure that TorchDIVA will produce repeatably consistent output across repeated trials. One notable exception to this deterministic nature is the vocal tract module of DIVA. The vocal tract is responsible for producing auditory and somatosensory feedback to the system in response to the articulatory motor commands sent by the motor cortex module, as well as generating the audio signal output at the end of the simulation process. Somatosensory and auditory feedback generated by the vocal tract module contains a desired amount of random perturbation. To control for this randomness in the vocal tract simulation, this added perturbation was held constant across DIVA and TorchDIVA during the comparison. For example, the vocal tract module introduces some minor pitch variability and noise in the produced sound signal. Validation of the vocal tract module required restricting randomly generated parameters in both models to constant, consistent values. This ensures that the random perturbations do not contribute to the normalized MSE observed between the two models.

## Extending TorchDIVA using DiffWave

DiffWave is a diffusion probabilistic model for generative audio synthesis first published by Kong et. al [12]. DiffWave is a bidirectional convolutional network capable of rapid and high-quality speech synthesis. Zhang et. al demonstrated that the DiffWave architecture could be modified to enhance degraded speech in a denoising application [18]. The authors show that by replacing the mel-spectrum upsampler component with a custom deep convolutional network (CNN), DiffWave can estimate an original speech waveform when presented with degraded speech. We followed a similar supervised training process, using the deep CNN mel-spectrum upsampler to enhance the speech quality of the TorchDIVA model. In our case, we treat the human speech sample as the original target speech waveform, and the TorchDIVA model output as the degraded version of the same audio sample. As shown in (Fig 2), the original DiffWave upsampler is extracted for use as a reference upsampler for training the deep CNN upsampler. During the training process, the reference upsampler produces a reference conditioner using the human speech mel-spectrum as input. An altered conditioner is generated from the TorchDIVA speech mel-spectrum. The new CNN upsampler is then trained on these conditioners using mean absolute error (L1) loss. After this training process, the CNN upsampler is combined with the remaining DiffWave network components and used to enhance the speech outputs generated by TorchDIVA.

**Fig 2. DiffWave Supervised Training Process.** Top: Process for training in the original DiffWave model. Bottom: Modified DiffWave training, using a deep CNN upsampler to match the conditioner in DiffWave's reference upsampler.

Supervised training of the modified mel-spectrum upsampler required samples of real human speakers and the TorchDIVA equivalent output. Human speech samples were obtained from the publicly available Saarbruecken voice database (SVD) [19]. The SVD dataset contains healthy and pathological speech recordings from 259 men and women producing the vowels /a/, /i/, and /u/. For this project, 50 healthy speakers were selected at random from the overall dataset. For each speaker, the pitch and formants F1-F3 were extracted from the sustained-pitch speech samples using the open-source Parselmouth library in Python [20]. Extracted formants were used to define a DIVA speech production target for each speech sample. This process yielded approximately 500 speech targets. The TorchDIVA model was then used to produce each target. Since these were new targets not encountered by the TorchDIVA model before, four iterations of training on the new target were run. This training step is necessary to allow the TorchDIVA model to learn appropriate articulator movements for each production target. The output of the fifth TorchDIVA production of the new target was then saved as an audio file. This process resulted in a paired dataset of human speech samples and the TorchDIVA model's attempt to approximate them. The mel-spectrum upsampler of the modified DiffWave model was then trained on the human speech waveform and conditioned on the TorchDIVA speech production [18]. The modified DiffWave model was lastly used to generate speech using only the TorchDIVA mel-spectrum as input. Ten speech samples used for the final generation step were held out of the training set. Finally, to evaluate the quality of the generated speech we used objective quality metrics utilized in the speech enhancement literature [21]. Specifically, we measured the perceptual evaluation of speech quality (PESQ), predicted rating of speech distortion (CSIG), predicted rating of background distortion (CBAK), segmental signal-to-noise ratio (segSNR) and predicted rating of overall quality (COVL). Each human speech sample was treated as the reference signal and compared against both the original and DiffWave-enhanced TorchDIVA speech samples to evaluate the difference in speech quality.

# Results

Each deterministic TorchDIVA module produces output that closely matches its DIVA counterpart. Due to differences in the PyTorch and Matlab programming languages, a negligible difference in output exists in the Motor Cortex implementation between the two models. Deterministic submodules were tested and confirmed to produce outputs which agree with the original model when presented with the same inputs. Below we describe the validation results by module.

## Motor Cortex

The TorchDIVA model is capable of both producing a well-learned speech target and learning to produce a new speech target over time. When learning a new speech target, both implementations can produce intelligible speech after the first three iterations. In Table 1 we provide the average normalized RMSE observed during the training and well-trained evaluations for each of the five speech targets tested. Over the course of 20 repetitions of the training process, the maximum normalized RMSE observed was 0.09%. When producing well-trained speech targets, the maximum normalized MSE observed was 0.12%. In (Figs 3 and 4), we plot the MSE over time in both the training and well-trained speech target cases for a selected speech target. This plot of the normalized RMSE over time reveals that the level of error stabilizes as training continues. The source of this minor error is explored in the Discussion section.

**Table 1. Normalized Root-Mean-Square Error (RMSE) of TorchDIVA Motor Signal.**

| Production Label | Training | Trained | Maximum |
|---|---|---|---|
| happy | 0.08% | 0.11% | 0.12% |
| i | 0.06% | 0.08% | 0.09% |
| u | 0.06% | 0.08% | 0.09% |
| e | 0.06% | 0.08% | 0.09% |
| ae | 0.06% | 0.08% | 0.09% |
| example | 0.06% | N/A[a] | 0.09% |

Average normalized root-mean-square error (RMSE) in motor command output between TorchDIVA and DIVA over 20 repetitions of the listed speech production. RMSE was measured under both training and well-trained speech target conditions. The final column contains the maximum observed RMSE over the 20 repetitions. The absolute difference between the two signals is expressed as a percentage of the magnitude of the signal range.

[a]The production labeled "example" has no default pretrained forward motor program.

**Fig 3. Normalized RMSE of Motor Command during Training.** Normalized root mean-square error (RMSE) in motor command output of TorchDIVA vs DIVA over 20 repetitions during the training process with the speech target 'u'.

**Fig 4. Normalized RMSE of Motor Command after Training.** Normalized root mean-square error (RMSE) in motor command output of TorchDIVA vs DIVA over 20 repetitions with a trained speech target 'u'.

## Vocal Tract

Direct comparison of the audio signal produced by the two models revealed some perturbation. The spectrogram provided in (Fig 5) illustrates this difference in the audio obtained from the two models. The absolute difference between the two audio signals was calculated using Matlab. Using the Matlab audioread function, both audio files were read into Matlab matrices. These matrices represent the amplitude of the audio signal, which is equivalent to the loudness of the sound. The audio signals are normalized over the range [-1, 1]. The measured difference has a maximum amplitude of 2e-3 Hz. Given the normalization, this error represents a deviation of approximately 0.1%. To determine whether this difference is detectable by a human listener, auditory excitation patterns (AEPs) were obtained and compared. Samples from DIVA and TorchDIVA were scaled to have the exact same loudness at 25.804 sones. The loudness and AEPs were measured using the ISO 532-2 method (Moore-Glasberg). Comparison of the AEPs revealed that the difference in the two audio signals is not perceivable by a human listener, as the AEP difference does not exceed Zwicker's 1 dB threshold at any frequency (Fig 6). This implies that the two samples are perceptually indistinguishable [22].

**Fig 5. DIVA and TorchDIVA Spectrogram Comparison.** Speech production 'happy' output audio comparison. The first subplot is DIVA, the second is TorchDIVA, and the bottom is the difference calculated from the two output signals.

**Fig 6. DIVA and TorchDIVA Auditory Excitation Pattern Comparison.** Speech production 'happy' auditory excitation pattern (AEP) comparison for DIVA and TorchDIVA. The first subplot is the AEP, the second subplot is the difference between the two AEPs obtained.

## TorchDIVA and DiffWave

Comparison of speech quality metrics confirms the DiffWave model can be trained to improve speech quality in the TorchDIVA model. The TorchDIVA model by default yields audio which does not closely match the human speakers. This result is expected given the varied speakers and the static vocal tract model used by TorchDIVA. In practice, the DIVA model (TorchDIVA or otherwise) struggles to produce speech sounds which fall outside of a narrow range of valid formants F1-F3. This limitation is reflected in generally low scores for the TorchDIVA output across all speech metrics used for the speech quality comparison (PESQ, CSIG, CBAK, COVL, segSNR). Speech samples generated by the modified DiffWave model reflect an improvement under all metrics evaluated (Fig 7a-e). The greatest improvement was seen in the CSIG metric. The mean CSIG score across the TorchDIVA speech samples was approximately -5.5. The mean CSIG score of the enhanced speech samples was approximately -1.3, reflecting a positive change of 4.2. These results provide one demonstration of the benefit that TorchDIVA offers over the original implementation; the modified DiffWave model can be readily integrated with it in order to enhance the speech quality.

**Fig 7. TorchDIVA and DiffWave Speech Quality Metrics.** Speech quality metric comparison between TorchDIVA and DiffWave-enhanced TorchDIVA samples. Original human speech sample is reference signal for all metric calculation. H_DIVA: human reference vs. TorchDIVA output. H_DW: human reference vs. DiffWave-enhanced output. **a)** Perceptual evaluation of speech quality (PESQ). **b)** Predicted rating of speech distortion (CSIG). **c)** Predicted rating of background distortion (CBAK). **d)** Predicted rating of overall quality (COVL). **e)** Segmental signal-to-noise ratio (segSNR).

## Discussion

PyTorch is used in many modern machine learning applications and provides a foundation which is flexible for further model enhancement. The TorchDIVA model has equivalent capabilities to the original and provides a valuable basis for future expansions. Testing shows that every DIVA module produces outputs which either precisely match the corresponding DIVA module or differ slightly as a result of numerical precision differences. For both well-trained and newly acquired speech targets, TorchDIVA can produce motor commands and audio which are perceptually indistinguishable from the original model.

Due to minor differences in the vocal tract calculation, there is a very small mismatch when directly comparing the two models. However, this difference has no discernible effect on the speech produced by the TorchDIVA model. The source of this mismatch comes from differences in the implementations of core computation routines in Python vs. Matlab. In DIVA, the routine *diva_synth_sample*, calculates the auditory and somatosensory result of a given motor command. The output of *diva_synth_sample* includes the fundamental frequency and formants (F0-F3), as well as the place of articulation, pressure, and voicing parameters of the vocal tract. Calculation of these parameters is based on a statically defined forward fit of the vocal tract model. One step of this calculation is a sum across a multidimensional array. Our testing revealed that when provided the same input, the PyTorch and Numpy equivalent sum operations produce an output which varies by a small degree relative to the Matlab version. Further investigation revealed that this minor error is due to numerical precision differences between the two programming languages. The *diva_synth_sample* routine is invoked in several places throughout the vocal tract module. The method is also used when calculating the Jacobian pseudoinverse in the inverse-map elements for feedback control. Furthermore, the DIVA model continually adjusts the synaptic weights of the feedforward motor program in response to the feedback control loop. As a result, these small variations introduced by *diva_synth_sample* cause a ripple effect through the TorchDIVA model. This variation is responsible for the minor motor command and auditory signal differences observed between the two models.

The TorchDIVA model is a ripe basis for further enhancement and optimization. One future goal of the TorchDIVA model is to replace the sequences of manual arithmetic operations taken in each module with small and efficient PyTorch neural networks. All TorchDIVA operations are performed using PyTorch operators and tensors, meaning that the PyTorch autograd engine can be used to train these networks with minimal additional development. This change has potential to accelerate the computational performance of the model. This is a significant advantage of the TorchDIVA model, as this process would be difficult or impossible to accomplish directly in Matlab Simulink. As demonstrated with DiffWave, TorchDIVA can be easily integrated with open-source speech tools and neural network models available in the Python ecosystem.

Novel speech-based machine learning algorithms can be developed based on DIVA. For example, the DIVA model is meant to accurately capture the process of human speech production, yet speech produced by the DIVA model is noticeably artificial to a human listener. We have shown that more realistic speech sounds can be produced by combining the DIVA model with DiffWave, an open-source neural vocoder built in PyTorch [17]. Re-creating DIVA in the PyTorch framework is an important move towards future DIVA enhancements of this nature. The modified DiffWave model enables TorchDIVA to produce more natural-sounding audio. However, the number of speech samples used for training and testing the modified DiffWave upsampler was low (492 train, 10 test). Speech quality yielded by the deep CNN upsampler may be improved by increasing the size of these datasets. By backpropagating the changes introduced by the generative DiffWave model into the TorchDIVA learning loop, one might enable the TorchDIVA model to produce higher quality speech natively. In addition, DIVA and TorchDIVA utilize a modified Maeda synthesizer of speech to generate speech sounds. Other articulatory vocal tract model options exist, some of which can approximate the speech dynamics of a real human speaker of a specific age or gender [23]. The TorchDIVA model can be integrated with these models to investigate whether dynamically adjusting these vocal tract dimensions can yield results which better approximate human speakers.

The TorchDIVA model has several limitations related to usability relative to the original implementation. TorchDIVA supports a basic command-line interface only, which may present a hurdle to those researchers who are accustomed to the graphical user-interface (GUI) of the original Matlab Simulink DIVA model. Additionally, TorchDIVA does not have a convenient way of defining speech targets manually by hand. Since new targets are defined in a plain-text format, this represents a minor limitation only. Users have the option of using DIVA or any text editor to generate new speech targets. Users also have the option to generate speech targets for TorchDIVA based on human speech recordings. Many avenues are available for further enhancement of TorchDIVA as a motor speech research platform. By providing the TorchDIVA source code to the wider research community, our aim is to empower researchers to modify the base implementation to suit their needs.

## Conclusion

TorchDIVA increases the utility and accessibility of the DIVA model for researchers in the speech neuroscience and speech machine learning (ML) community. TorchDIVA enables continual expansion and refinement of DIVA through integrations with deep learning models, data visualization packages, and more sophisticated vocoder models. This has the potential to further extend the utility of DIVA as a research tool in speech neuroscience and to serve as a theoretically grounded constraint for machine learning model development in the speech ML community.

# Acknowledgements

This work was partially funded by grant NIH-NIDCD R01DC006859.

# Data Availability Statement

The source code for TorchDIVA can be found at:

https://github.com/skinahan/DIVA_PyTorch

**Figures**

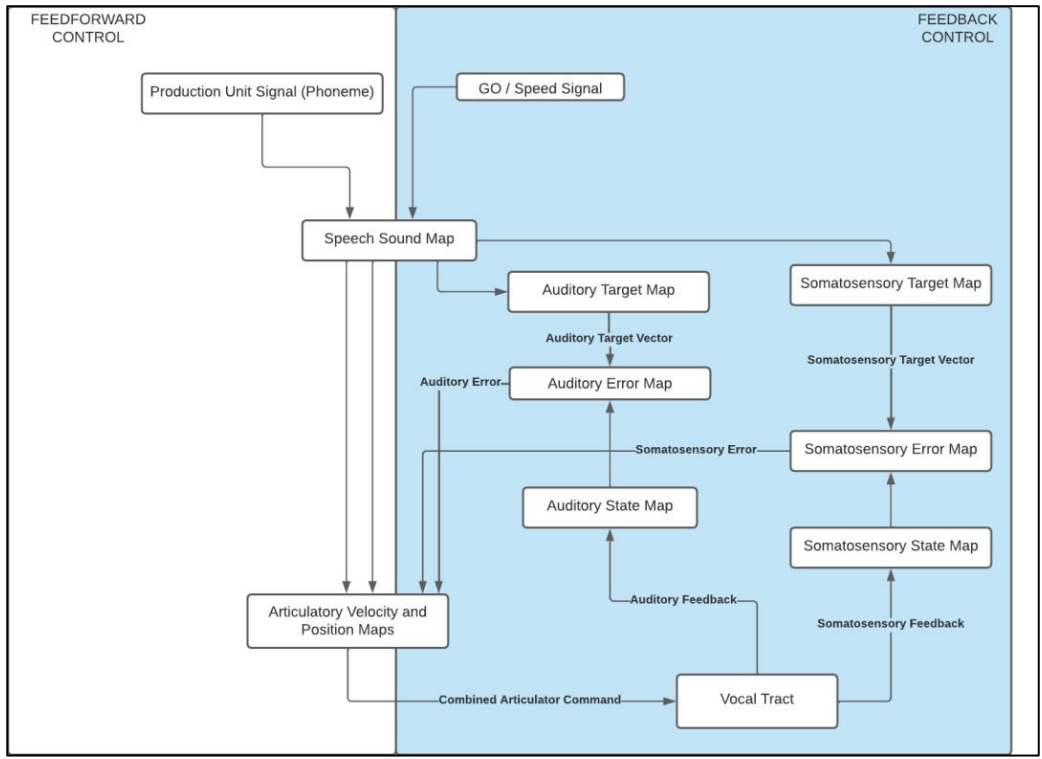

**Fig 1. DIVA Model Architecture.** Simplified schematic view of the DIVA model, showing the combination of feedforward and feedback control loops.

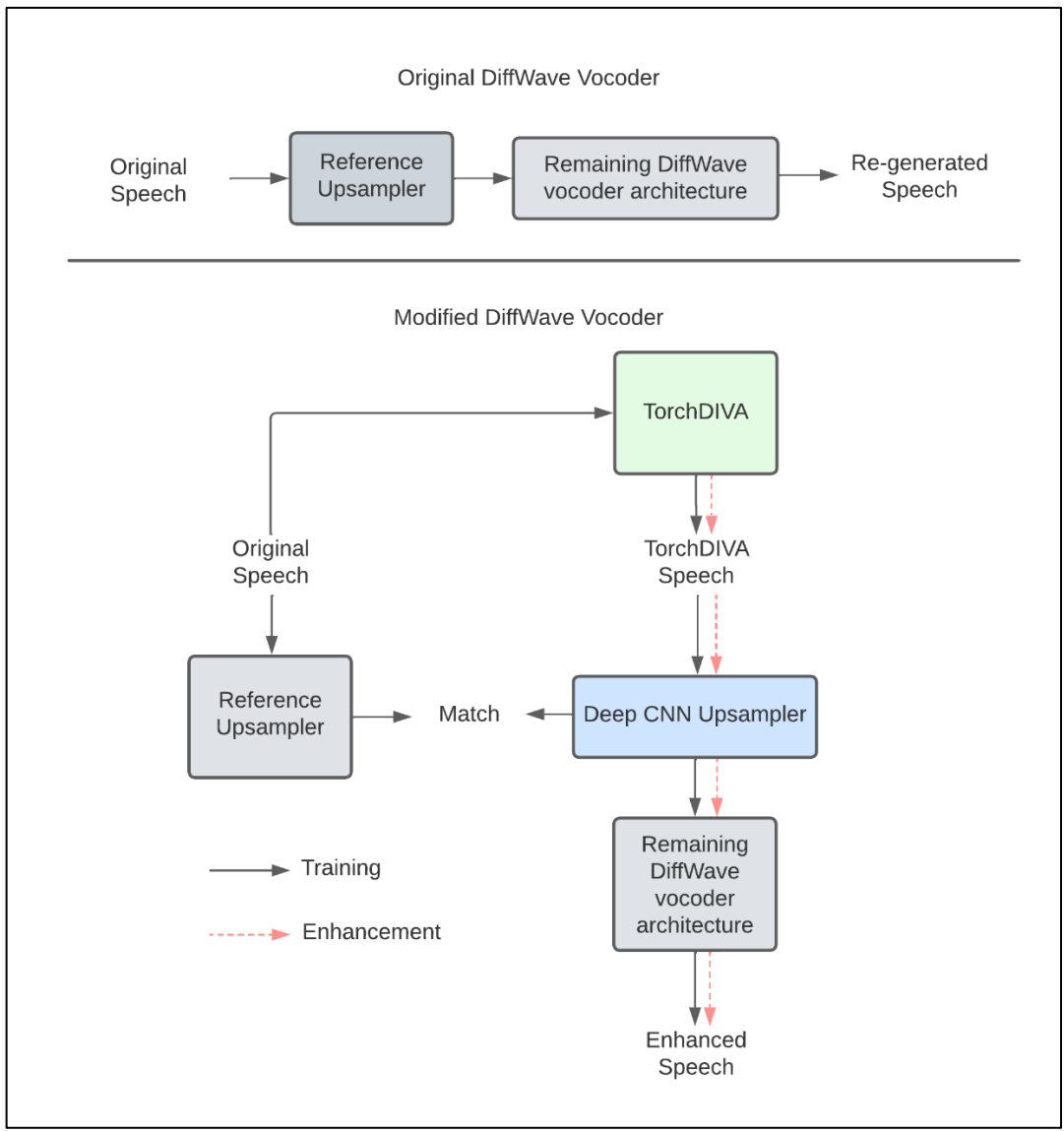

**Fig 2. DiffWave Supervised Training Process.** Top: Process for training in the original DiffWave model. Bottom: Modified DiffWave training, using a deep CNN upsampler to match the conditioner in DiffWave's reference upsampler.

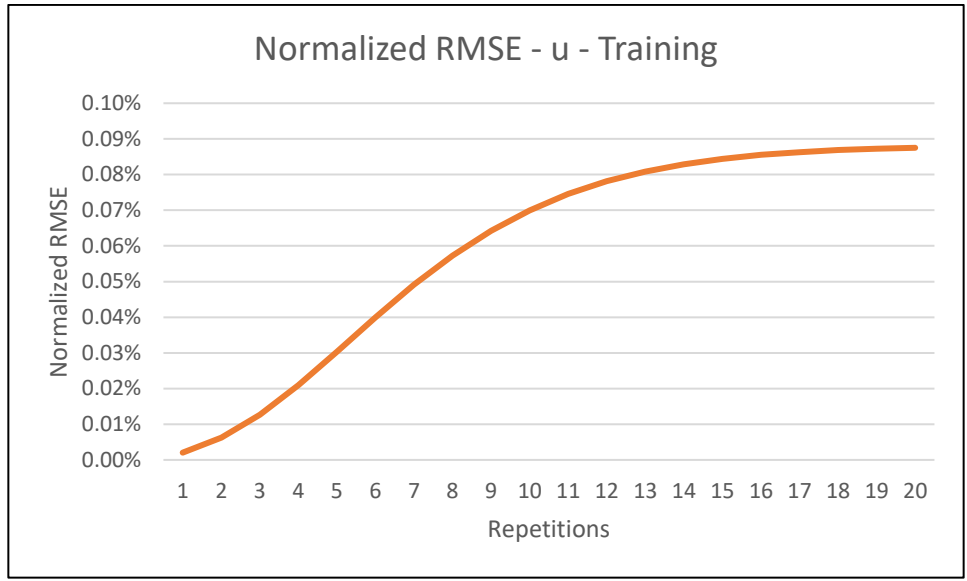

**Fig 3. Normalized RMSE of Motor Command during Training.** Normalized root mean-square error (RMSE) in motor command output of TorchDIVA vs DIVA over 20 repetitions during the training process with the speech target 'u'.

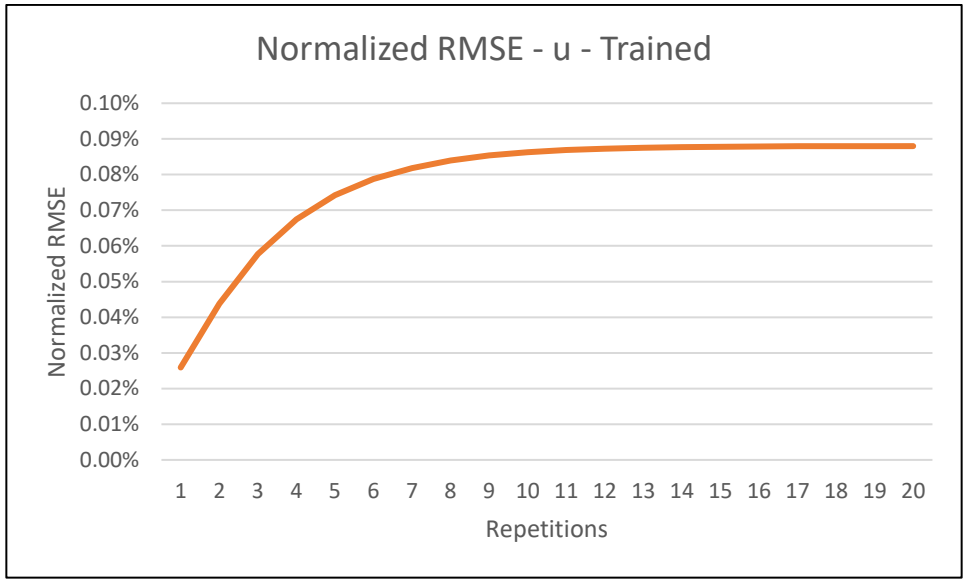

**Fig 4. Normalized RMSE of Motor Command after Training.** Normalized root mean-square error (RMSE) in motor command output of TorchDIVA vs DIVA over 20 repetitions with a trained speech target 'u'.

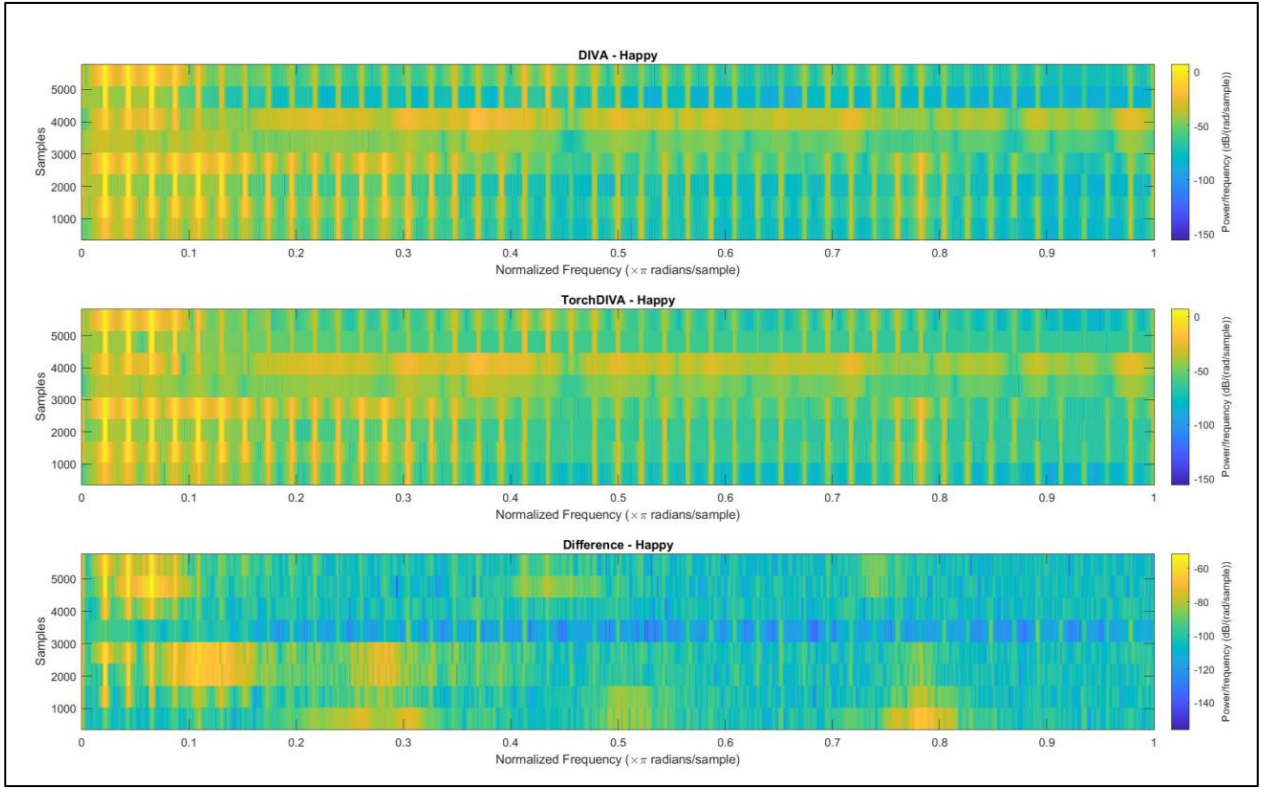

**Fig 5. DIVA and TorchDIVA Spectrogram Comparison.** Speech production 'happy' output audio comparison. The first subplot is DIVA, the second is TorchDIVA, and the bottom is the difference calculated from the two output signals.

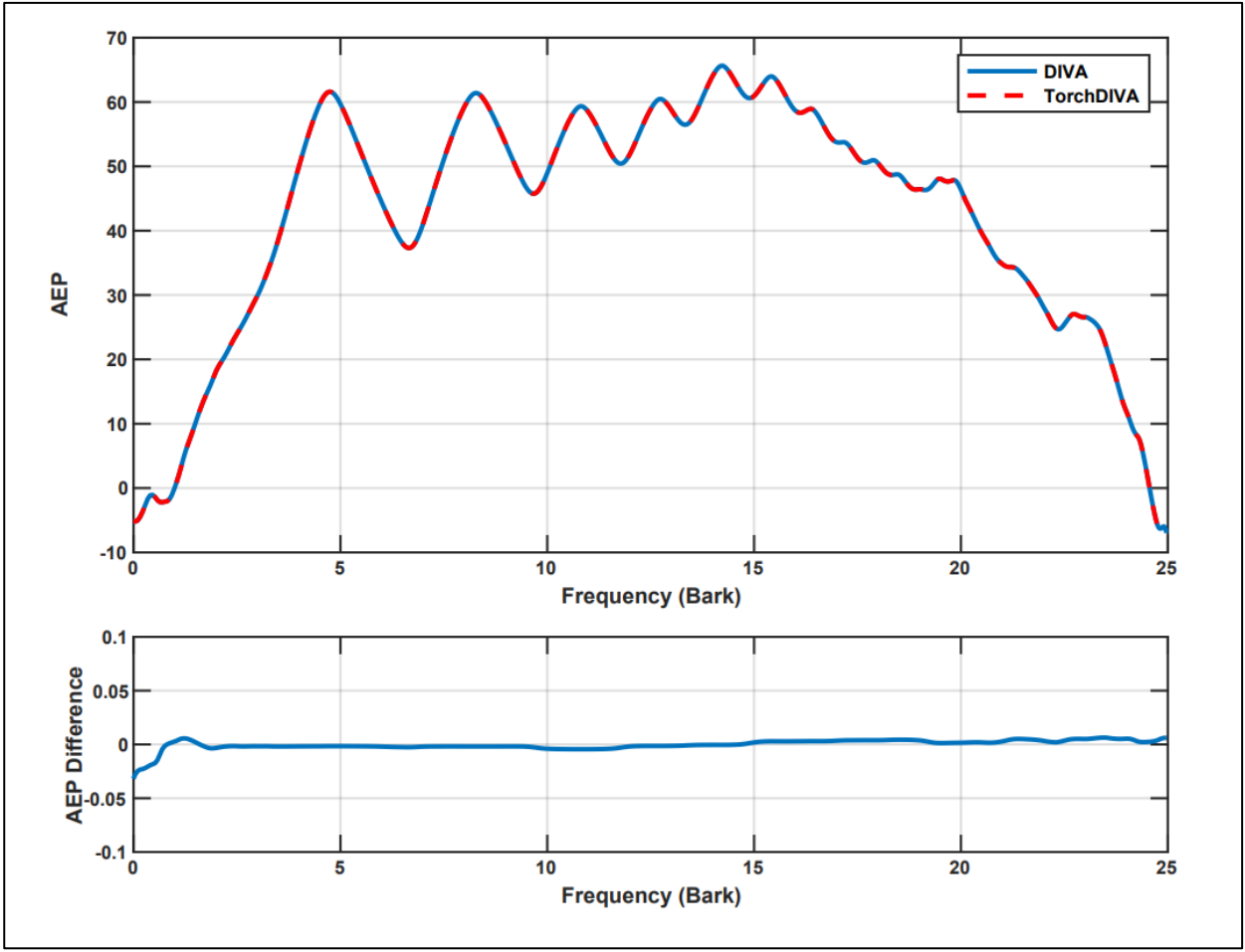

**Fig 6. DIVA and TorchDIVA Auditory Excitation Pattern Comparison.** Speech production 'happy' auditory excitation pattern (AEP) comparison for DIVA and TorchDIVA. The first subplot is the AEP, the second subplot is the difference between the two AEPs obtained.

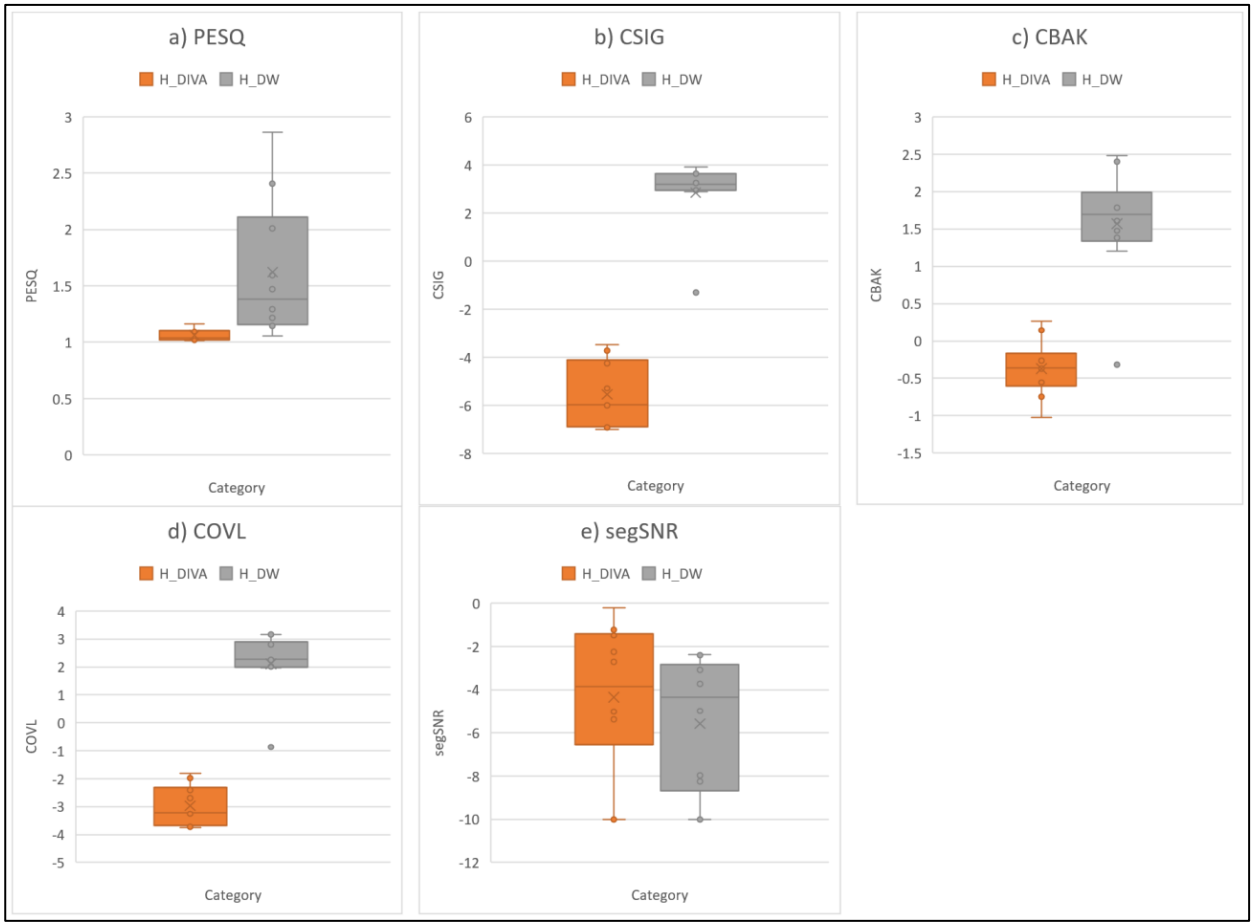

**Fig 7. Comparison of TorchDIVA and DiffWave Speech Quality Metrics.** Speech quality metric comparison between TorchDIVA and DiffWave-enhanced TorchDIVA samples. Original human speech sample is reference signal for all metric calculation. H_DIVA: human reference vs. TorchDIVA output. H_DW: human reference vs. DiffWave-enhanced output. **a)** Perceptual evaluation of speech quality (PESQ). **b)** Predicted rating of speech distortion (CSIG). **c)** Predicted rating of background distortion (CBAK). **d)** Predicted rating of overall quality (COVL). **e)** Segmental signal-to-noise ratio (segSNR).